# Evolution of as-a-Service Era in Cloud

**Sugam Sharma**

Center for Survey Statistics and Methodology, Iowa State University, Ames, Iowa, USA

Email: sugamsha@iastate.edu

**Abstract.** Today, a paradigm shift is being observed in science, where the focus is gradually shifting toward the cloud environments to obtain appropriate, robust and affordable services to deal with Big Data challenges (Sharma et al. 2014, 2015a, 2015b). Cloud computing avoids any need to locally maintain the overly scaled computing infrastructure that include not only dedicated space, but the expensive hardware and software also. In this paper, we study the evolution of as-a-Service modalities, stimulated by cloud computing, and explore the most complete inventory of new members beyond traditional cloud computing stack.

## 1. Introduction

Today, the cloud computing also has emerged as one of the major shifts in recent information and communication age that promises an affordable and robust computational architecture for large-scale and even for overly complex enterprise applications (Fenn et al. 2008). It is a powerful and revolutionary paradigm that offers service-oriented computing and abstracts the software-equipped hardware infrastructure from the clients or users. Although, the concept of cloud computing is mainly popular in three praxises- 1) IaaS (Infrastructure-as-a-Service ), 2) PaaS (Platform-as-a-Service ), and 3) SaaS (Software-as-a-Service ), but in this data science age, should be equally expandable to DBaaS (Database-as-a-Service ) (Curino, et al., 2011; Seibold et al., 2012). Over the last few year, the cloud computing has evolved as scalable, secure and cost-effective solutions to an overwhelming number of applications from diversified areas. The traditional cloud ecosystem delivers the services in three popular flavors -IaaS, PaaS, and SaaS. The scientific communities understand the purview of pervasiveness of cloud computing and intent to fan out its technological advantages to benefit every single field of study. The imagination of ubiquitous information access, free from the geography constraints, and supported by the distinguish characteristics of cloud computing has encouraged the communities enough to further extend the cloud horizon beyond IaaS, PaaS, and SaaS. This gives rise to the accelerated evolution of "as-a-Service" (aaS) framework in almost all the domains. Specifically, highlighted is an extended list of new entrants of aaS family to the cloud computing stack on the top of IaaS, PaaS, and SaaS.

 Rest of the paper is structured as follows. Section 2 highlights the evolution of as-a-Service era and mentions the most complete list of new entrant of aaS family. Finally, the paper is concluded in section 3.

## 2. Evolution of as-a-Service Era

Cloud computing is considered as one of most accomplished and ubiquitous paradigms in 21$^{st}$ century, especially for the service related computing. It is a financially economical and technologically robust invention that has revolutionized the delivery of IT resources and services into as-a-Service framework.

### 2.1. Traditional cloud services

In this section, a general classification of some of the leading cloud services is provided (Figure 3). Each class is briefly described here.

1. **Software-as-a-Service (SaaS).** SaaS is one of the most popular, repository-rich and widely used cloud model that is offering the services for more than a decade now. The service repository is much diversified and cover a very wide range of simple to complex services such as Google Email, Google Doc, etc. Under the contractual agreement of SaaS model, the vendor also called the service provider is fully responsible to provide all the essential infrastructure consisting of the robust hardware resources and expansive software systems. Also provided is a graphical user interface (GUI) that facilitates the user interaction with the service.

2. **Infrastructure-as-a-Service (IaaS).** As, the name suggests, IaaS offers the infrastructure as service. The infrastructure consists of various building blocks, which can be combined or layered to derive a customized environment most appropriate to execute the designated applications. Some of the most popular examples of IaaS cloud model include Amazon Web Services (AWS), Rackspace, etc.

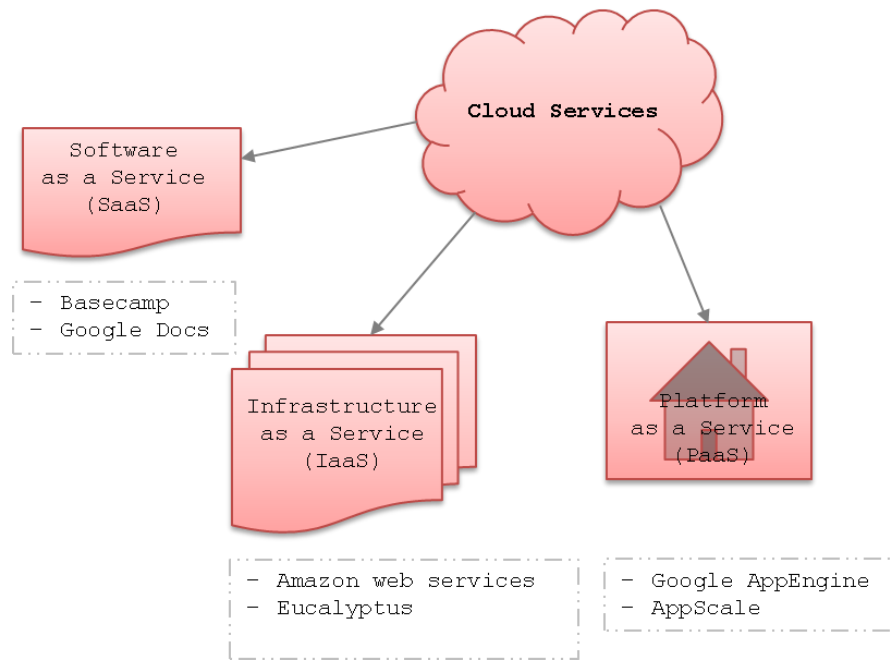

Figure 3. Traditional cloud services

3. **Platform-as-a-Service (PaaS).** In the PaaS cloud model, the service provider is responsible to provide the risk free and robust environment for software product development. The environment consists of required software tools and is hosted on the hardware infrastructure of the service provider. A user can avail the benefits of the PaaS for the software development, either by using the APIs provided or thorough a GUI. Google App Engine, Acquia.com, Force.com, etc. are some of the popular example of PaaS.

**2.2 Emerging cloud services**

1. **Beaconing-as-a-Service (BaaS)** is a service-oriented beaconing strategy for vehicular ad hoc networks (Lasowski et al., 2012).

2. **Biometric Authentication-as-a-Service (BioAaaS)** is a biometric technologies powered authentication approach to perceived privacy and data protection risks in web environments, which is based on the SaaS (Senk et al., 2011).

3. **Business Integration-as-a-Service (BIaaS/BIaS)** enables connections between services operating in the Cloud and integrates different services and business activities to achieve a streamline process (Chang et al., 2012).

4. **Business Intelligence-as-a-Service (SaaS BI/BIaaS)** is a new cloud computing paradigm for Business Intelligence processes that offers data access through a web interface. The implementations and details are hidden from the users. Consequently, the business processes are orchestrated in a simpler and faster manner (Sano, 2014).

5. **Cashier-as-a-Service (CaaS)** refers to those merchant websites that accept payments through third-party cashiers such as PayPal, Amazon Payments and Google Checkout (Wang et al., 2011).

6. **Climate Analytics-as-a-Service (CAaaS)** is a solution to address Big Data challenges in climate science. CAaaS combines high-performance computing and data-proximal analytics with scalable data management, cloud computing virtualization, the notion of adaptive analytics, and a domain-harmonized API to improve the accessibility and usability of large collections of climate data (Schnase, et al., 2015).

7. **Cloud-Based Analytics-as-a-Service (CLAaaS)** is a platform for big data analytics, which provides on demand data storage and analytics services through customized user interfaces that include query, decision management, and workflow design and execution services for different user groups (Zulkernine et al., 2013).

8. **Confidentiality-as-a-Service (CaaS)** is a paradigm to provide usable confidentiality and integrity for the bulk of users, for whom the current security mechanisms are too complex or require too much effort. It focuses on unobtrusive confidentiality by hiding all cryptographic artefacts from the prevalently non-technical users. Data protection is based on symmetric encryption and invisible key-management mechanisms (Fahl et al., 2012).

9. **Content Distribution-as-a-Service (CoDaas)** is enables on demand virtual content delivery service overlays for user-generated content providers to deliver the contents to a group of designated consumers. Built on a hybrid media cloud, it offers an elastic private virtual content delivery service with an agreed quality of service (Jin et al., 2012).

10. **Continuous Analytics-as-a-Service (CaaaS)** is a cloud computing model for enabling convenient, on demand network access to a shared pool of continuous analytics results of real-time events such as monitoring oil & gas production, watching traffic status and detecting accident (Chen et al., 2011).

11. **Cooperation-as-a-Service (CaaS)** is a new service-oriented solution that enables improved and new services for the road users and an optimized use of the road network through vehicle's cooperation and vehicle-to-vehicle communications (Mousannif et al., 2012).

12. **Crimeware-as-a-Service (CaaS)** is a business model used in the underground market where illegal services are provided to help underground buyers conduct cybercrimes (such as attacks, infections, and money laundering) in an automated manner. It provides a new dimension to cybercrime by making it more organized, automated, and accessible to criminals with limited technical skills (Sood et al., 2013).

13. **Data-as-a-Service (DaaS)** is a type of cloud-assisted services that deliver data on demand to consumers through APIs. DaaS helps the consumers to avoid the typical need to fetch and store giant data assets and then search for the required information in the data asset (Vu et al., 2012).

14. **Data Integrity-as-a-Service (DIaaS)** enables all the expertise related to data integrity in one place to deal with some of the well-known problems such as public verifiability and dynamic content. DIaaS not only releases the burdens of data integrity management from a storage service by handling it through an independent third party data Integrity Management Service (IMS), but also reduces the security risk of the data stored in the storage services by checking the data integrity with the help of IMS (Nepal et al., 2011).

15. **Data Mining-as-a-Service (DMAS/DMaaS)** allows the data owners to leverage hardware and software solutions provided by DMAS providers, without developing their own. It is especially for those clients, who have large volume of data but limited budget for data analysis, to outsource their data and data mining needs to a third-party service provider. (Liu et al., 2012).

16. **Database-as-a-Service (DBaaS/DaaS)** promises to move much of the operational burden of provisioning, configuration, scaling, performance tuning, backup, privacy, and access control from the database users to the service operator, offering lower overall costs to users. Early DBaaS examples include Amazon RDS and Microsoft SQL Azure that promise to address the market need for such a service (Curino, 2011). The DBaaS cloud model is gaining popularity in this data science age (Wolfe, 2013). The service agreement under DBaaS assures the shift of data management related responsibilities, especially administration from end user to a third party vendor, called service provider, who can be public or private. Some of the data related operational burdens include upgrade, provisioning, failover management, configuration, seamless scaling, performance tuning, backup, privacy, access control (Seibold and Kemper, 2012), etc. DBaaS not only migrates the responsibilities, but it guarantees overall a very lower costs to the end users. Also, on the client site, the DBaaS model avoids any needs of a professionally trained database administrator (DBA) who is primarily responsible of data management and maintenance otherwise. In DBaaS, the full access to a complex database can be achieved through very simple service calls only. The users, who use these services, do not feel that they are interacting with any database. Therefore, the engineers, engaged in application development are neither expected nor required to have expert understanding of database management. The data related operations such as failover, scaling, etc. in DBaaS cloud is expected not to impact any live user in any mean. In, sum, a DBaaS offers - 1) a shared, consolidated platform to provision database services on, 2) a self-service model for provisioning those resources, 3) elasticity to scale out and scale back database resources, 4) chargeback based on database usage. Some of the early DBaaS providers include Amazon RDS and Microsoft SQL Azure.

17. **DDoS-as-a-Service (DDoSaaS)** now empowers a naïve user, not having advanced knowledge, with the ability to launch Denial of Service (DDoS) attacks and is offering it-as-a-Service, open to everybody. DDoS attacks are an increasing threat on the Internet. On contrast, previously, until couple of years ago, the users were required to be knowledge-rich in computer networks, consequently, launches of such attacks were limited (Santanna et al., 2014).

18. **Description-as-a-Service (DESCaaS)** aims to extend the architecture of information systems by adding resource description functionality. It provides the uniform descriptions of resources and their content to improve data accessibility, interoperability, and discovery. The offered description, allows users to publish, find, and access distributed resources efficiently (Henry et al., 2012).

19. **Desktop-as-a-Service (DaaS)** has drawn considerable attention in recent years due to its formation of a significant interface over the gap between clustered servers and clients. DaaS paradigm complies with the client-server model that makes the information stored in servers on network and cached on clients. However, most implementations rely on continuous display synchronization between user interface on the client and application logic on the server through various protocols on data transmissions (Shu et al., 2012).

20. **Digital Forensics-as-a-Service (DFaaS)** is a new service-based approach for processing and investigating the high volume of seized digital material. It aids reducing the case backlogs and freeing up digital investigators to help detectives better understand the digital material. Now a days, this approach has become a standard for hundreds of criminal cases and over a thousand detectives (Baar et al., 2014).

21. **Digital Intellectual Property Resources-as-a-Service (DIPaaS)** access on a flexible economic basis is a major interest among the users today. DIP is the human intellectual work in digital form and eBook, software program, e-painting, movie, song, and computer game are some of the examples. The services engaged in DIP are complex and encompass creator, manufacturer, distributor, licensing agencies, and service providers as important components. DIPaaS offers the associated services to have access to DIP on demand and avoids all the above stated intermediary components and satisfies the users, especially mobile at large (Mohiuddin et al., 2013).

22. **Disaster Tolerance-as-a-Service (DTaaS)** is a cloud-based service for disaster tolerance. It extends the Remus virtualization-based high availability system by allowing groups of virtual machines to be replicated across data centers over wide-area Internet links. SecondSite is one example of DTaaS (Rajagopalan et al., 2012).

23. **Education and learning-as-a-Service (ELaaS)** can be understood as the use of cloud computing in the educational and learning environment. ElaaS enables the learners, instructors, and administrators to perform their tasks effectively by using the services it offers that results in overall education cost reduction (Alabbadi, 2011).

24. **Energy-as-a-Service (EaaS)** is a concept that uses multimedia cloud computing to ease off the limitations on energy capacity of smartphones, which have experienced a dramatic growth in number. EaaS is able to address the challenges of limited energy capacity in an efficient way by offloading heavy tasks to the cloud (Altamimi et al., 2012).

25. **Everything-as-a-Service (XaaS)** also called as anything-as-a-service, facilities the flexibility for users and companies to all on demand customize their computing environments to craft the desired experiences. The components that XaaS is highly dependent on are 1) a strong cloud services platform, 2) reliable Internet connectivity to successfully gain traction and acceptance among both individuals and enterprises (Duan, 2012).

26. **Exploits-as-a-Service (EaaS)** is a paradigm that is very effective in malware ecosystem. It allows the attackers pay for an exploit kit or service to do the "dirty work" of exploiting a victim's browser, decoupling the complexities of browser and plugin vulnerabilities from the challenges of generating traffic to a website under the attacker's control. After a successful exploit, these kits load and execute a binary provided by the attacker, effectively transferring control of a victim's machine to the attacker (Grier et al., 2012).

27. **Failure-as-a-Service (FaaS)** allows the cloud services to routinely perform the large-scale failure drills in real deployments. FaaS enables cloud services to routinely exercise large-scale failures online, which will strengthen individual, organizational, and cultural ability to anticipate, mitigate, respond to, and recover from failures. Before it experiences any unexpected failure scenarios, a cloud service could perform failure drills from time to time to find out the real-deployment scenarios in which its recovery does not work (Gunawi et al., 2011).

28. **Failure scenario-as-a-Service (FSaaS)** will be utilized across the cloud for testing the resilience of cloud applications. FSaaS provides both Hadoop service and application vendors a way to test their applications against the risk of massive failure (Faghri et al., 2012).

29. **Fault Masking-as-a-Service (FAS/FMaaS)** is an external web service, which offers the composite services to register at will. Subsequently, post registration, it periodically checks the partner links, detects unavailable partner services, and updates the composite service with available alternatives (Gülcü et al., 2014).

30. **Financial Modeling and Prediction-as-a-Service (FMPaaS)** can be understood as the financial domain being powered by robustness of cloud computing. The related operations are delivered to customers/consumers as services. The accuracy and performance are the key benefits of such services, which are achieved through FMPaaS (Chang et al., 2015).

31. **Forensics-as-a-Service (FRaaS)** provides a comprehensive cloud forensics solution for creating a repeatable system. Such a system could be implemented as a standard forensics operational model for deployment within the cloud ecosystem regardless of environments and client service lines (Shende et al., 2014).

32. **Gaming-as-a-Service (GaaS)** is the upcoming trend in the game industry. Similar to cloud computing services in other domains, the cloud gaming services are gaining popularity and exhibit several advantages over the traditional software systems. Scalability, ubiquitous and cross-platform support, and cost-effectiveness over the terminal hardware constraints are some of the example features of GaaS (Cai et al., 2014).

33. **Hadoop-as-a-Service (HDaaS)** enables the enterprises to perform various Hadoop-based operations such as analytics, management, and storage of Big Data in a cloud in a cost-effective and time-efficient manner. Hence, eliminates the need for any on premise hardware. The Hadoop architecture and the supporting applications are abstracted into a single cloud-based delivery. HDaaS platform is provided as a web-based subscription service on a pay-per-use basis. The HDaaS platform enables enterprises to use the Hadoop technology in a cost-effective manner, while ensuring minimal time consumption (Wood, L. (2015).

34. **Handwritten Character Recognition-as-a-Service (HCRaaS)** is a cloud-based recognition platform that is useful for large scale character recognition, writing adaptation technology, and handwriting Chinese word/text line recognition that usually involve large storage and computation cost. With HCRaaS, the mobile devices are no longer subject to local computing capacity and storage resource constraints. It also delivers the higher recognition accuracy and personalized service with low hardware costs and can provide reliable handwriting solution across different mobile OS with higher recognition performance (Gao et al., 2011).

35. **Hardware-as-a-Service (HaaS)** a new cloud layer in the existing traditional cloud stack of IaaS, PaaS, and SaaS. It allows the usage of distinct hardware components through the Internet similar to the cloud services. The remote hardware, distributed over multiple geographical locations is transparently integrated into an operating system. This gives a feel as if all hardware devices are locally connected to the local system (Stanik et al., 2012).

36. **HPC-as-a-Service (HPCaaS)** is a new approach that uses a cloud abstraction to provide a simple interface to high-end HPC resources. The users can pay for HPC resources as needed. The HPCaaS framework transforms a supercomputer into an elastic cloud of multiple federated clouds that supports dynamic provisioning, efficient utilization, and maximum accessibility of HPC resources through IaaS, PaaS, and SaaS abstractions (AbdelBaky et al., 2012).

37. **Identity-as-a-Service (IdnaaS)** is one of most promising models that deliver identify management through cloud computing in the field of e-Government. The cloud service providers currently implement IdnaaS by through a central identity broker that serves as a hub between various services and identity providers (Zwattendorfer et al., 2013).

38. **Intrusion Detection-as-a-Service (IDaaS)** is a cloud powered framework for an intrusion detection and reporting service for consumers based on the type of application and consumer's security needs (Veigas et al., 2013).

39. **Laboratories-as-a-Service (LaaS)** is a model for developing remote laboratories as independent component modules and implementing them as a set of loosely-coupled services to be consumed with a high level of abstraction and virtualization. It tackles the common concurrent challenges in remote laboratories developing and implementation such as inter-institutional sharing, interoperability with other heterogeneous systems, coupling with heterogeneous services and learning objects, difficulty of developing, and standardization (Tawfik et al., 2013).

40. **Manufacturing-as-a-Service (MFGaaS)** also known as cloud manufacturing is a computing and service oriented manufacturing model developed from existing advanced manufacturing models enterprise information technologies under the support of cloud computing, IoT, virtualization and service-oriented technologies, and advanced computing technologies. It aims to realize the full sharing and circulation, high utilization, and on-demand use of various manufacturing resources and capabilities by providing safe and reliable, high quality, cheap and on-demand used manufacturing services for the whole lifecycle of manufacturing (Tao et al., 2013).

41. **Mobility-as-a-Service (MobiaaS)** provides the consumers the required connectivity service continuity and seamless handover for flows like voice as the consumers use a multitude of devices to communicate. This requires a range of different heterogeneous networks, specific connectivity services and different mobility approaches for individual. It also ensures that the consumers are always reachable and have consistent and personalized services such as the location awareness and network capabilities (Baliga et al., 2011).

42. **Mobility Prediction-as-a-Service (MPaaS)** embeds mobility mining and forecasting algorithms into a cloud-based user location tracking framework. MPaaS drives new fashion of mobile cloud applications and it equips a hybrid predictor fusing to provide telecom cloud with large-scale mobility prediction capacity (Xiong et al., 2014).

43. **Monitoring-as-a-Service (MaaS)** helps users to deploy state monitoring at different levels of cloud services compared with developing ad hoc monitoring tools or setting up dedicated monitoring infrastructure. It brings cloud service providers the opportunity to consolidate monitoring demands at different levels (infrastructure, platform, and application) to achieve efficient and scalable monitoring and leverages the state-of-the-art monitoring tools and functionalities (Meng et al., 2013).

44. **Object-as-a-Service (ObaaS)** is based on the notion of building dynamically the service needed on each object and then integrate it in the whole composition. ObaaS runs on the Object, using its functionalities such as sensing, actuating, and computing. A specific service is dynamically created and tailored for user's need, on-the-fly, following the description given by the user on the programming layer offered by the object (Cherrier et al., 2014).

45. **Ontology-as-a-Service (OaaS)** is an ontology tailoring process in the cloud, which underneath exercises the sub-ontology extraction and replacement on the cloud (Flahive et al., 2014).

46. **Policing-as-a-Service (PolaaS)** is offered by the cloud providers with the intention of empowering users to monitor and guard their assets in the cloud. A cloud provider is able to offer basic auditing services due to undeveloped tools and applications. A user can purchase the desired service out of the available pool to gain some control over their data (Zargari et al., 2014).

47. **Policy Management-as-a-Service (IPMaaS)** a cloud based policy management framework that is designed to give users a unified control point for managing access policies to control access to his resources no matter where they are stored (Takabi et al., 2012).

48. **Proximity-as-a-Service (ProxaaS)** is a concept, where any exiting or even especially created Wi-Fi hot spot could be used as presence sensor that can trigger access for some user-generated information snippets. According to ProxaaS helps discover the hyper local data as information snippets that are relevant for mobile subscribers being at this moment nearby some Wi-Fi access point (Namiot et al., 2012).

49. **RAN-as-a-Service (RANaaS)** delivers a flexible architecture based on centralized processing. It is capable of handling the increasing interference in very dense networks. As a result, it helps reducing the energy consumption. Thereby, it aids in cost-efficiently deploying and managing cellular networks (Sabella et al., 2013).

50. **Risk-Assessment-as-a-Service (RAaaS)** is a new paradigm that serves as an autonomic method to measure the risk, especially the security-related, of the cloud. It is based the on-demand, automated, multi-tenant architecture of the cloud. It gets the continuous "risk score" of the cloud environment with respect to a given tenant or any specific application. That more generally is useful to new tenants and applications (Kaliski et al., 2010).

51. **Routing-as-a-Service (RaaS)** is a framework that is developed for tenant-directed route control in data centers. The solutions that are based on RaaS are operational even on the commercial, off-the-shelf hardware running legacy technologies. This indicates that the solution implementation is possible on the existing networks without major infrastructural overhaul (Chen et al., 2011).

52. **Secure Logging-as-a-Service (SecLaaS)** assists storing the virtual machines' logs. Also, it grants the access to forensic investigators to ensure the confidentiality of the cloud users. Furthermore, SeclaaS preserves the proofs of previous log and this helps in protecting the integrity of the logs from any fraudulent or dishonest cloud providers or investigators (Zawoad et al., 2013).

53. **Security-as-a-Service (SecuaaS/SaaS)** is a data protection and a host & application protection solution. It validates the security services over a geographically distributed, large scale, multi-cloud and federated cloud infrastructure (Pawar et al., 2015).

54. **Sensing and Actuation-as-a-Service (SAaaS)** is a step forward to create a cloud of sensors and actuators. As, the cloud provides on-demand computing and storage of the resources with guaranteed quality of service; moving the sensors and actuators, that will be able to interact with the surrounding environment, will enable the development of new and value added services and opens avenue for pervasive cloud computing (Distefano et al., 2012).

55. **Sensing-as-a-Service (S²aaS)** provides the cloud-powered sensing services through mobile phones to various domains such as environmental monitoring, social networking, healthcare, transportation, etc. S²aaS is an energy-efficient cloud system and has the ability to support various mobile phone sensing applications on different smartphone platforms (Sheng et al., 2013).

56. **Smart City-as-a-Service (SCaaS)** is a cloud-enabled platform that is deployed as a vehicle for catalyzing smart city innovation. The emergence of global government clouds (G-cloud) has exhibited the propitious potential to enable the development of smart cities. SCaaS aims to capitalize on G-cloud and the nascent innovation capabilities of other technological paradigm to facilitate the smart city development (Clohessy et al., 2014).

57. **Social Context-as-a-Service (SoCaaS/SCaaS)** is platform that helps managing the adaptations of collaborative pervasive applications. The objective of SCaaS is to facilitate the active interactions among a dynamic groups that may contain users, stakeholders, businesses, etc., dubbed as actors. The characteristics of the relationships between actors are based upon the predefined agreements and constraints. The modeling process of a relationship has the notion of social context (Kabir et al., 2014).

58. **Software Development-as-a-Service (SDaaS)** is being offered by IBM Almaden Research Center. The SDaaS requires to have a good understanding of the business and pain points of the clients. The collected knowledge then helps to build an application. SDaaS aims to at least meet the customer's need, but may exceeds their expectations in an ideal scenario (Lehman et al., 2011).

59. **Storage-as-a-Service (StaaS/SaaS)** provides an online storage space in cloud to store the data. The users' concerns about the data security and privacy are addressed by robust cryptographic algorithms in place. StaaS also rewards the users the optimized computation cost, higher security level and sensitivity based on significance of the data (Patel et al., 2012).

60. **Supply Chain-as-a-Service (SCaaS)** is a cloud powered supply chain systems. It makes feasible the supply chains to be offered as the set of services. The consumer's demand are received as service requests. The coordination to determine the optimal service compositions becomes sometime challenging (Leukel et al., 2011).

61. **Test-bed-as-a-Service (TBaaS/TaaS)** provides a cloud-based ready-to-go environment/ infrastructure for the experimental activities. It also provides easy access to the required communications, computing and storage resources for the experiments (Aragó et al., 2014).

62. **Testing-as-a-Service (TeaaS/TaaS)** is a new business and service model in cloud infrastructure that undertakes the software testing related project activities or jobs for a under-test web-based software system, which are delivered to customers as the services. It is a flexible model that accelerates the quality implementation of the product. In sum, it delivers the on-demand test execution of well-defined suites of test material, generally on outsourced basis (Gao et al., 2013).

63. **Things-as-a-Service (ThiaaS)** is a way where the innovative as well as the value-added services are implemented through the integration of the cloud computing with the Internet of Things. It is a way to develop a cloud of things where the heterogeneous resources are aggregated or abstracted based on the tailored thing-like semantics (Distefano et al., 2012).

64. **Threat-as-a-Service (ThraaS/TaaS)** assists the malicious actors (users) to equip them with the ability to find out and overwhelm the vulnerable targets with heavy traffic. The geographic locale or organizations are not discriminated by the felonious users. ThraaS can be enabled on demand from any compromised location, infrastructure, or even from the attacker's house (Tsai et al., 2011).

65. **Ticketing-as-a-Service (TicaaS/TaaS)** is a cloud based low cost flexible ticketing system, where the ticketing-related core processes are leveraged on a Software-as-a-Service (SaaS) model. TicaaS has migrated the business logic from the standalone terminal to pay-per-use cloud. The ticketing terminal equipment such as validators, vending machines, etc. are bundled in the cloud environment. The scalability in electronic ticketing is adequately addressed by the elasticity of the cloud (Ferreira et al., 2013).

66. **Trust-as-a-Service (TraaS/TaaS)** is a framework that ameliorates the management of trust feedbacks in cloud environment. An adaptive credibility relies on cloud service consumers' capability and majority consensus of their feedbacks to distinguish between credible trust and malicious feedbacks (Noor et al., 2011).

67. **Variability-as-a-Service (VaaS)** model accommodates the outsourced burden of some of the IT resources from traditional SaaS (Software-as-a-Service). Variability management in various customer's requirements during the multi-tenant modeling of a single application instance by SaaS is the key concern that is offered-as-a-Service by VaaS. It induces the appearance of VaaS providers also (Ghaddar et al., 2012).

68. **Virtual cluster-as-a-Service (ViteraaS)** supplies the need of an on demand high performance computing, especially for the complex projects engaging longer research simulations, and widely operated e-Learning and teaching in a privately owned cloud. ViteraaS is flexible and customizable and is able to integrate with public cloud infrastructure IaaS, PaaS or within the university's existing IT infrastructure like Single Sign-On for seamless authentication and authorization, if required. ViteraaS keeps track of the performance and status of associated virtual machines and QoS (Quality of Service) monitoring module to harvest the required information (Doelitzscher et al., 2011).

# 3 Conclusions

The thrust of this work was to study the impact of cloud computing motivation in the evolution of as-a Service era. The most complete list of the new members of as-a-Service family was explored. The list comprised several new as-a-Service platforms far beyond the traditional cloud computing stack that typically consisted of IaaS, PaaS, and SaaS.